\documentclass[12pt]{iopart}
\usepackage{amssymb}
\usepackage{amsmath}
\usepackage{amsfonts}
\usepackage{iopams}

\usepackage[numbers]{natbib}
\usepackage{url}
\newcommand{\m}{\mathbf{m}}

\newcommand{\deriv}[2]{\frac{d #1}{d #2}}

\renewcommand{\S}{{\mathbf \m}}

\newcommand{\diag}[1]{\,{\sf diag}\left( #1 \right)}

\usepackage{xspace}

\newcommand{\bfA}{\mathbf{A}}
\newcommand{\bfB}{\mathbf{B}}

\newcommand{\bfD}{\mathbf{D}}

\newcommand{\bfH}{\mathbf{H}}

\newcommand{\bfJ}{\mathbf{J}}

\newcommand{\bfP}{\mathbf{P}}

\newcommand{\bfW}{\mathbf{W}}

\newcommand{\bff}{\mathbf{f}}

\newcommand{\bfk}{\mathbf{k}}

\newcommand{\bfm}{\mathbf{m}}

\newcommand{\bfv}{\mathbf{v}}
\newcommand{\bfw}{\mathbf{w}}

\newcommand{\bfy}{\mathbf{y}}
\newcommand{\bfz}{\mathbf{z}}

\newcommand{\bftheta}{\boldsymbol{\theta}}
\newcommand{\bfpsi}{\boldsymbol{\psi}}
\newcommand{\bfPsi}{\boldsymbol{\Psi}}

\newcommand{\nn}{^{n+1}}
\newcommand{\nnm}{^{n+1,m}}
\newcommand{\nnmm}{^{n+1,m+1}}
\newcommand{\n}{^{n}}

\newcommand{\bfdo}{\mathbf{d}_{\rm obs}}
\newcommand{\bfdp}{\mathbf{d}_{\rm pred}}

\newcommand{\FF}{F}
\usepackage{graphicx}   

\usepackage{lineno}

\begin{document}

\title{A numerical method for efficient 3D inversions using Richards equation}

\author{R Cockett$^1$, L J Heagy$^1$, E Haber$^1$}

\address{$^1$ Department of Earth, Ocean and Atmospheric Sciences, 2020-2207 Main Mall, Vancouver, BC Canada V6T 1Z4}
\ead{rowanc1@gmail.com}
\vspace{10pt}
\begin{indented}
\item[]\today
\end{indented}

\begin{abstract}
Fluid flow in the vadose zone is governed by Richards equation; it is parameterized by hydraulic conductivity, which is a nonlinear function of pressure head. Investigations in the vadose zone typically require characterizing distributed hydraulic properties. Saturation or pressure head data may include direct measurements made from boreholes. Increasingly, proxy measurements from hydrogeophysics are being used to supply more spatially and temporally dense data sets. Inferring hydraulic parameters from such datasets requires the ability to efficiently solve and deterministically optimize the nonlinear time domain Richards equation. This is particularly important as the number of parameters to be estimated in a vadose zone inversion continues to grow. In this paper, we describe an efficient technique to invert for distributed hydraulic properties in 1D, 2D, and 3D. Our algorithm does not store the Jacobian, but rather computes the product with a vector, which allows the size of the inversion problem to become much larger than methods such as finite difference or automatic differentiation; which are constrained by computation and memory, respectively. We show our algorithm in practice for a 3D inversion of saturated hydraulic conductivity using saturation data through time. The code to run our examples is open source and the algorithm presented allows this inversion process to run on modest computational resources.
\end{abstract}

%
\submitto{\IP}
\vspace{2pc}
\vspace{2pc}
%
\maketitle
%
%

\section{Introduction}

Studying the processes that occur in the vadose zone, the region between the earth's surface and the fully saturated zone, is of critical importance for understanding our  groundwater resources. Fluid flow in the vadose zone is described by Richards equation and parameterized by hydraulic conductivity, which is a nonlinear function of pressure head \cite{Richards1931, Celia1990}. Typically, hydraulic conductivity is heterogeneous and can have a large dynamic range. In any site characterization, the spatial estimation of the hydraulic conductivity function is an important step. Achieving this, however, requires the ability to efficiently solve and optimize the nonlinear, time-domain Richards equation. Rather than working with a full, implicit, 3D time-domain system of equations, simplifications are consistently used to avert the conceptual, practical, and computational difficulties inherent in the parameterization and inversion of Richards equation. These simplifications typically parameterize the conductivity and assume that it is a simple function in space, often adopting a homogeneous or one dimensional layered soil profile (cf. \citep{Binley2002, Deiana2007, Hinnell2010, Liang2014}). Due to the lack of constraining hydrologic data, such assumptions are often satisfactory for fitting observed measurements, especially in two and three dimensions as well as in time. However, as more data become available, through spatially extensive surveys and time-lapse proxy measurements (e.g. direct current resistivity surveys and distributed temperature sensing), extracting more information about subsurface hydrogeologic parameters becomes a possibility. The proxy data can be directly incorporated through an empirical relation (e.g. \citep{Archie1942}) or time-lapse estimations can be structurally incorporated through some sort of regularization technique \citep{HaberHoltzman2013, ho, Hinnell2010}. Recent advances have been made for the forward simulation of Richards equation in a computationally-scalable manner \citep{RichardsFOAM}. However, the inverse problem is non-trivial, especially in three-dimensions \citep{Towara2015}, and must be considered using modern numerical techniques that allow for spatial estimation of hydraulic parameters.

Inverse problems in space and time are often referred to as history matching problems (see \citep{DeanChen2011, OliverBook2008,  SarmaDurlofskyAziz2007, Oliver01, hydrusCalibration2012} and reference within). Inversions use a flow simulation model, combined with some a-priori information, in order to estimate a spatially variable hydraulic conductivity function that approximately yields the observed data. The literature shows a variety of approaches for this inverse problem, including trial-and-error, stochastic methods, and various gradient based methods \citep{Bitterlich2004, Binley2002, Carrick2010, Durner1994, Finsterle2011c, Mualem1976, Simunek1996}. The way in which the computational complexity of the inverse method scales becomes important as problem size increases \citep{Towara2015}. Computational memory and time often become a bottleneck for solving the inverse problem, both when the problem is solved in 2D and, particularly, when it is solved in 3D \citep{hao}. To solve the inverse problem, stochastic methods are often employed, which have an advantage in that they can examine the full parameter space and give insights into non-uniqueness \citep{Finsterle2011}. However, as the number of parameters we seek to recover in an inversion increases, these stochastic methods require that the forward problem be solved many times, which often makes these methods impractical. This scalability, especially in the context of hydrogeophysics has been explicitly noted in the literature (cf. \cite{Binley2002, Deiana2007, Towara2015, Linde2016}).

Derivative-based optimization techniques become a practical alternative when the forward problem is computationally expensive or when there are many parameters to estimate (i.e. thousands to millions). Inverse problems are ill-posed and thus to pose a solvable optimization problem, an appropriate regularization is combined with a measure of the data misfit to state a deterministic optimization problem \citep{tikhonov1977}. Alternatively, if prior information can be formulated using a statistical framework, we can use Bayesian techniques to obtain an estimator through the Maximum A Posteriori model (MAP) \citep{somersallo}. In the context of Bayesian estimation, gradient based methods are also important, as they can be used to efficiently sample the posterior \citep{BuiThanhGhattas2015}.

A number of authors have sought solutions for the inverse problem, where the forward problem is Richards equation (cf. \citep{Bitterlich2002, Iden2007, hydrusCalibration2012} and references within). Since the problem is parabolic (therefore, stiff), most work discretizes the forward problem by an implicit method in time and a finite volume or finite element in space. Most work uses a Newton-like method for the resulting nonlinear system, which arises from the discretization of the forward problem. For the deterministic inverse problem using Richards equation, previous work uses some version of a Gauss-Newton method (e.g. Levenberg-Marquardt), with a {\bf direct} calculation of the sensitivity matrix \citep{Finsterle2011, Simunek1996, Bitterlich2002}. However, while these approaches allow for inversions of moderate scale, they have one major drawback: the sensitivity matrix is large and dense; its computation requires dense linear algebra and a non-trivial amount of memory (cf. \citep{Towara2015}). Previous work used either external numerical differentiation (e.g. PEST) or automatic differentiation in order to directly compute the sensitivity matrix \citep{Finsterle2011c, Bitterlich2002, Doherty2004, Towara2015}. These techniques can generate inaccuracies in the sensitivity matrix and, consequently, tarry the convergence of the optimization algorithm. Furthermore, external numerical differentiation is computationally intensive and limits the number of model parameters that can be estimated.

The goal of this paper is to suggest a modern numerical formulation that allows the inverse problem to be solved {\bf without explicit} computation of the sensitivity matrix by using {\bf exact} derivatives of the discrete formulation \citep{hao}. Our technique is based on the discretize-then-optimize approach, which discretizes the forward problem first and then uses a deterministic optimization algorithm to solve the inverse problem \cite{Gunz03}. To this end, we require the discretization of the forward problem. Similar to the work of \citep{Celia1990}, we use an implicit Euler method in time and finite volume in space. Given the discrete form, we show that we can analytically compute the derivatives of the forward problem with respect to distributed hydraulic parameters and, as a result, obtain an implicit formula for the sensitivity. The formula involves the solution of a linear time-dependent problem; we avoid computing and storing the sensitivity matrix directly and, rather, suggest a method to efficiently compute the product of the sensitivity matrix and its adjoint times a vector. Equipped with this formulation, we can use a standard inexact Gauss-Newton method to solve the inverse problem for distributed hydraulic parameters in 3D. This large-scale distributed parameter estimation becomes computationally tractable with the technique presented in this paper and can be employed with any iterative Gauss-Newton-like optimization technique.

This paper is structured as follows: in Section~\ref{sec:richards-forward}, we discuss the discretization of the forward problem on a staggered mesh in space and backward Euler in time; in Section~\ref{sec:richards-inverse}, we formulate the inverse problem and construct the implicit functions used for computations of the Jacobian-vector product. In Section~\ref{sec:richards-validation}, we demonstrate the validity of the implementation of the forward problem and sensitivity calculation. Section~\ref{sec:richards-iterative} compares two iterative approaches, Picard and Newton, for handling the nonlinearity in Richards equation. Finally, in Section~\ref{sec:richards-examples}, we show an example of a 3D inversion for hydraulic conductivity and discuss extensions for inverting for multiple distributed hydraulic parameters from Richards equation and contrast the scalability of our methodology to standard techniques.

To accelerate both the development and dissemination of this approach, we have built these tools on top of an open source framework for organizing simulation and inverse problems in geophysics \citep{simpeg2015}. We have released our numerical implementation under the permissive MIT license. Our implementation of the implicit sensitivity calculation for Richards equation and associated inversion implementation is provided and tested to support 1D, 2D, and 3D forward and inverse simulations with respect to custom empirical relations and sensitivity to parameters within these functions. The source code can be found at \url{https://github.com/simpeg/simpeg} and may be a helpful resource for researchers looking to use or extend our implementation.

\section{Forward problem}
\label{sec:richards-forward}

In this section, we describe Richards equations and its discretization \citep{Richards1931}. Richards equation is a nonlinear parabolic partial differential equation (PDE) and we follow the so-called mixed formulation presented in \citep{Celia1990} with some modifications. In the derivation of the discretization, we give special attention to the details used to efficiently calculate the effect of the sensitivity on a vector, which is needed in any derivative based optimization algorithm.
\subsection{Richards equation}

The parameters that control groundwater flow depend on the effective saturation of the media, which leads to a nonlinear problem. The groundwater flow equation has a diffusion term and an advection term which is related to gravity and only acts in the $z$-direction. There are two different forms of Richards equation; they differ in how they deal with the nonlinearity in the time-stepping term. Here, we use the most fundamental form, referred to as the `mixed'-form of Richards equation \citep{Celia1990}:

\begin{equation}
\label{eq:richards-mixed}
    \frac{\partial \theta(\psi)}{\partial t} - \nabla \cdot k(\psi) \nabla \psi - \frac{\partial k(\psi)}{\partial z} = 0
    \quad \psi \in \Omega
\end{equation}
where $\psi$ is pressure head, $\theta(\psi)$ is volumetric water content, and $k(\psi)$ is hydraulic conductivity. This formulation of Richards equation is called the `mixed'-form because the equation is parameterized in $\psi$ but the time-stepping is in terms of $\theta$. The hydraulic conductivity, $k(\psi)$, is a heterogeneous and potentially anisotropic function that is assumed to be known when solving the forward problem. In this paper, we assume that $k$ is isotropic, but the extension to anisotropy is straightforward \citep{simpeg2015, fvtutorial}. The equation is solved in a domain, $\Omega$, equipped with boundary conditions on $\partial \Omega$ and initial conditions, which are problem-dependent.


An important aspect of unsaturated flow is noticing that both water content, $\theta$, and hydraulic conductivity, $k$, are functions of pressure head, $\psi$. There are many empirical relations used to relate these parameters, including the Brooks-Corey model \citep{Brooks1964} and the van Genuchten-Mualem model \citep{Mualem1976, VanGenuchten1980}. The van Genuchten model is written as:

\begin{subequations}
\label{eq:van-genuchten}
\begin{equation}
\label{eq:van-genuchten-water-retention}
    \theta(\psi) =
    \left\{\begin{aligned}
        \theta_r& + \frac{\theta_s- \theta_r}{(1+|\alpha \psi|^n)^m}  & \psi < 0 \\
        \theta_s& & \psi \ge 0
    \end{aligned}\right.
\end{equation}
\begin{equation}
\label{eq:van-genuchten-hydraulic-conductivity}
    k(\psi) =
    \left\{\begin{aligned}
        K_s & \theta_e(\psi)^l(1-(1- \theta_e(\psi)^{-m})^m)^2 & \psi < 0 \\
        K_s& & \psi \ge 0
    \end{aligned}\right.
\end{equation}
\end{subequations}
where
\begin{equation}
\label{eq:van-genuchten-params}
    \theta_e(\psi) = \frac{\theta(\psi) - \theta_r}{\theta_s - \theta_r},
    \qquad
    m=1- \frac{1}{n},
    \qquad
    n > 1
\end{equation}

\begin{figure}[ht]
\begin{center}
\includegraphics[width=1.0\textwidth]{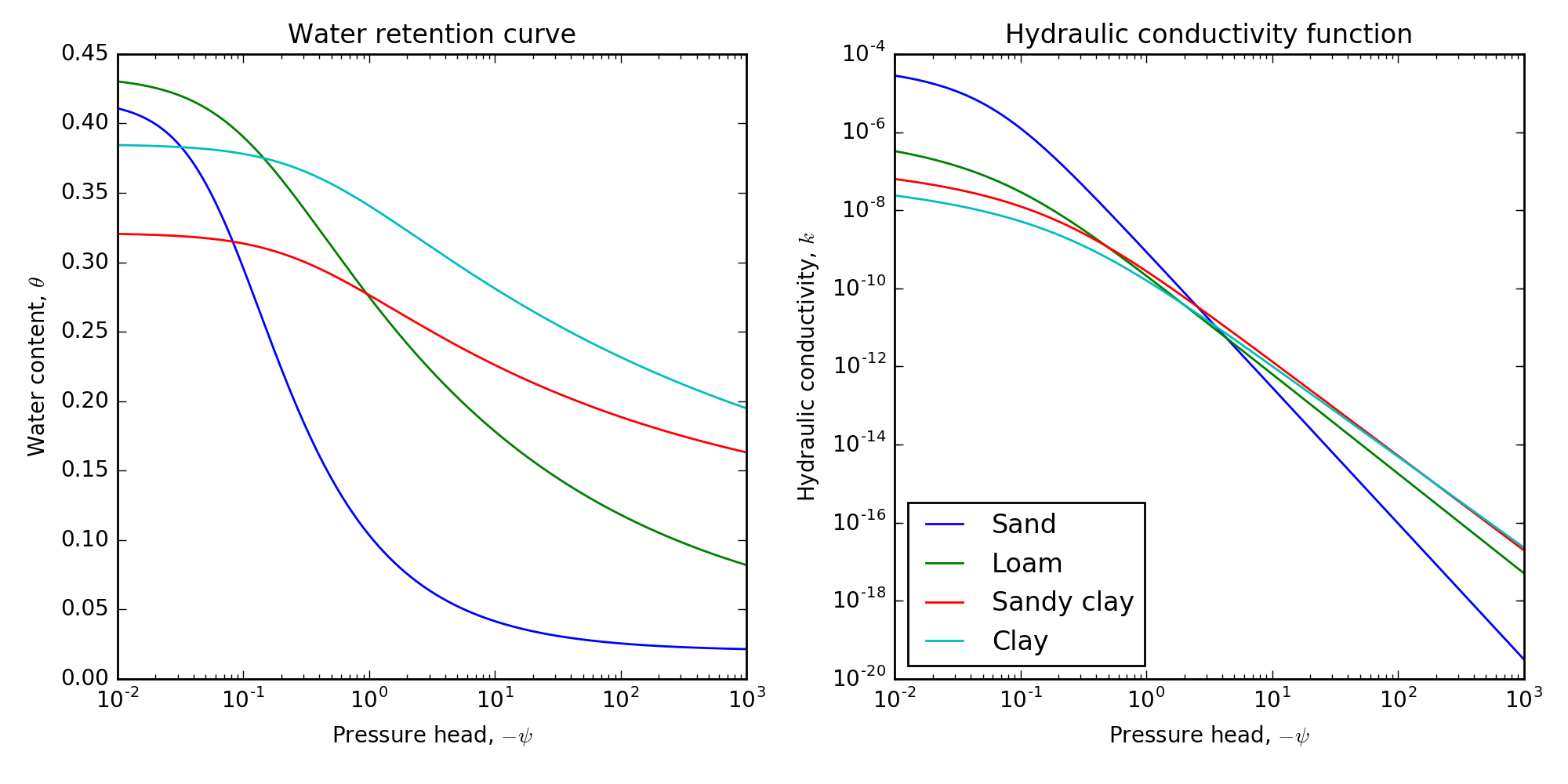}
\end{center}
\caption{The water retention curve and the hydraulic conductivity function for four canonical soil types of sand, loam, sandy clay, and clay.}
\label{fig:van-genuchten}
\end{figure}

Here, $\theta_r$ and $\theta_s$ are the residual and saturated water contents, $K_s$ is the saturated hydraulic conductivity, $\alpha$ and $n$ are fitting parameters, and, $\theta_e(\psi) \in [0,1]$ is the effective saturation. The pore connectivity parameter, $l$, is often taken to be $\frac{1}{2}$, as determined by \cite{Mualem1976}. Figure~\ref{fig:van-genuchten} shows the functions over a range of negative pressure head values for four soil types (sand, loam, sandy clay, and clay). The pressure head varies over the domain $\psi \in (-\infty, 0)$. When the value is close to zero (the left hand side), the soil behaves most like a saturated soil where $\theta = \theta_s$ and $k = K_s$. As the pressure head becomes more negative, the soil begins to dry, which the water retention curve shows as the function moving towards the residual water content ($\theta_r$). Small changes in pressure head can change the hydraulic conductivity by several orders of magnitude; as such, $k(\psi)$ is a highly nonlinear function, making Richards equation a nonlinear PDE.

\subsection{Discretization}
Richards equation is parameterized in terms of pressure head, $\psi$. Here, we describe simulating Richards equation in one, two, and three dimensions. We start by discretizing in space and then we discretize in time. This process yields a discrete, nonlinear system of equations; for its solution, we discuss a variation of Newton's method.

\subsubsection{Spatial Discretization}

In order to conservatively discretize Richards equation, we introduce the flux ${\vec f}$ and rewrite the equation as a first order system of the form:

\begin{subequations}
\label{eq:richards-mixed-first-order}
\begin{eqnarray}
\label{eq:richards-mixed-first-order-a}
&&    \frac{\partial \theta(\psi)}{\partial t} - \nabla \cdot {\vec f} - \frac{\partial k(\psi)}{\partial z} = 0 \\
\label{eq:richards-mixed-first-order-b}
&&    k(\psi)^{-1} {\vec f} =  \nabla \psi
\end{eqnarray}
\end{subequations}

\begin{figure}[!htbp]
\begin{center}
\includegraphics[width=0.9\textwidth]{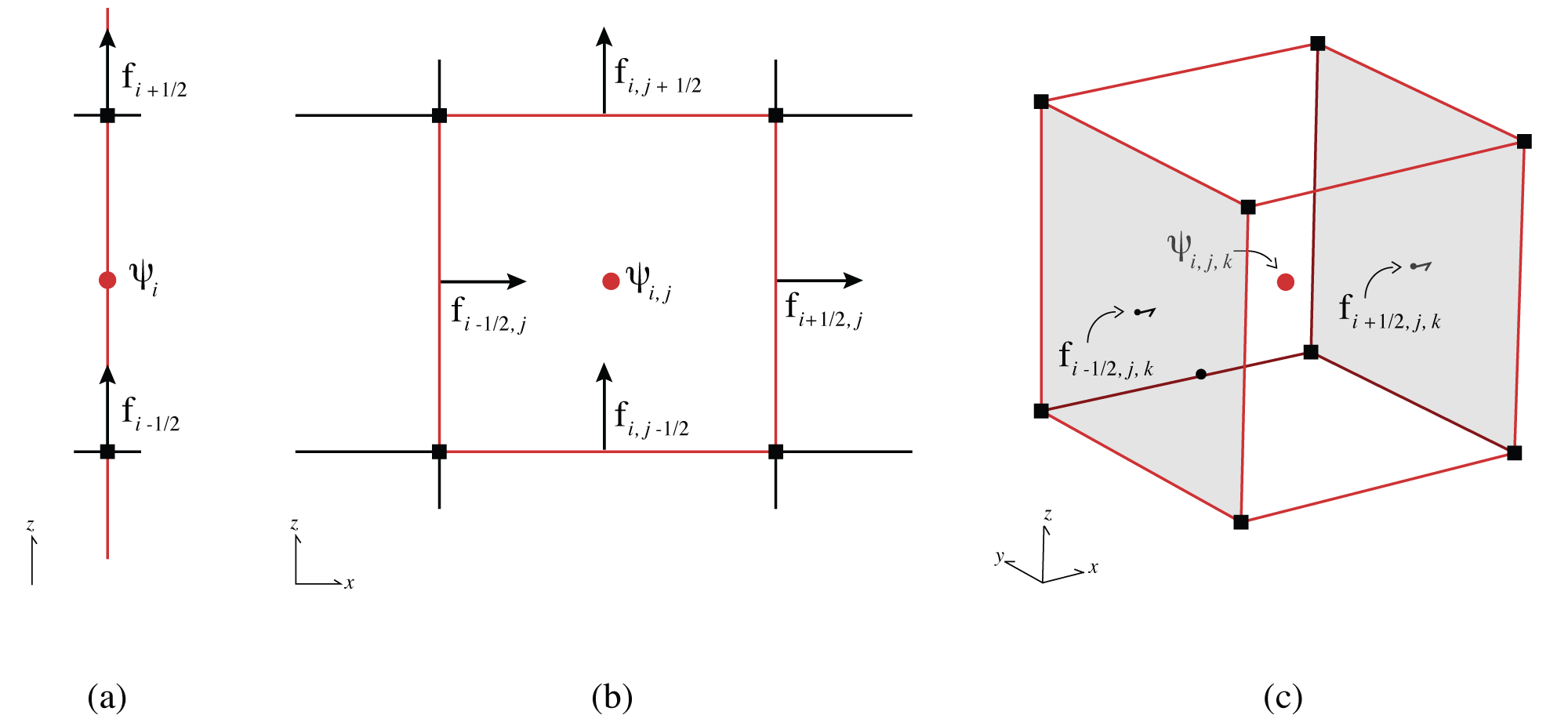}
\end{center}
\caption{Discretization of unknowns in 1D, 2D and 3D space. Red circles are the locations of the discrete hydraulic conductivity $K$ and the pressure head $\psi$.
The arrows are the locations of the discretized flux $\vec f$ on each cell face.}
\label{fig:richards-finite-volume}
\end{figure}

We then discretize the system using a standard staggered finite volume discretization (cf. \cite{Ascher2008, haber2015computational, fvtutorial}). This discretization is a natural extension of mass-conservation in a volume where the balance of fluxes into and out of a volume are conserved \citep{LipnikovMisitas2013}. Here, it is natural to assign the entire cell one hydraulic conductivity value, $k$, which is located at the cell center. Such assigning leads to a piecewise constant approximation for the hydraulic conductivity and allows for discontinuities between adjacent cells. From a geologic perspective, discontinuities are prevalent, as it is possible to have large differences in hydraulic properties between geologic layers in the ground. The pressure head, $\psi$, is also located at the cell centers and the fluxes are located on cell faces, which lead to the usual staggered mesh or Marker and Cell (MAC) discretization in space \citep{fletcher}. We demonstrate the discretization in 1D, 2D and 3D on the tensor mesh in Figure~\ref{fig:richards-finite-volume}. We discretize the function, $\psi$, on a cell-centered grid, which results in a grid function, $\bfpsi$. We use bold letters to indicate other grid functions.

The discretization of a diffusion-like equation on an orthogonal mesh is well-known (see \citep{Haber2001,fletcher,HaberHeldmannAscher07,Ascher2011} and reference within). We discretize the differential operators by using the usual mass balance consideration and the elimination of the flux, $\bff$ \footnote{Here we assume an isotropic conductivity that leads to a diagonal mass matrix and this yields easy elimination of the fluxes.}. This spatial discretization leads to the following discrete nonlinear system of ordinary differential equations (assuming homogeneous Dirichlet boundary conditions):

\begin{equation}
\label{eq:richards-mixed:discrete}
\frac{d \boldsymbol{\theta}(\boldsymbol{\psi})}{d t}
- \mathbf{D}
    \text{ diag}
    \left(
        \mathbf{k}_{Av}(\boldsymbol{\psi}^{n+1})
    \right)
\mathbf{G} \boldsymbol{\psi}
- \mathbf{G}_z
    \left(
        \mathbf{k}_{Av}(\boldsymbol{\psi}^{n+1})
    \right)
=0
\end{equation}
Here, $\bfD$ is the discrete divergence operator and $\mathbf{G}$ is the discrete gradient operator. The discrete derivative in the $z$-direction is written as $\mathbf{G}_z$. The values of $\psi$ and $k(\psi)$ are known on the cell-centers and must be averaged to the cell-faces, which we complete through harmonic averaging \citep{Haber2001}.

\begin{equation}
\label{eq:mass-matrix-kav}
\mathbf{k}_{Av}(\boldsymbol{\psi}) = \frac{1}{\mathbf{A}_v\frac{1}{\mathbf{k}(\boldsymbol{\psi})}}
\end{equation}
where $\bfA_{v}$ is a matrix that averages from cell-centers to faces and the division of the vector is done pointwise; that is, we use the vector notation, $(1/{\bfv})_{i} = 1/\bfv_{i}$. We incorporate boundary conditions using a ghost-point outside of the mesh \citep{tos}.

\subsubsection{Time discretization and stepping}

Richards equation is often used to simulate water infiltrating an initially dry soil. At early times in an infiltration experiment, the pressure head, $\psi$, can be close to discontinuous. These large changes in $\psi$ are also reflected in the nonlinear terms $k(\psi)$ and $\theta(\psi)$; as such, the initial conditions imposed require that an appropriate time discretization be chosen. Hydrogeologists are often interested in the complete evolutionary process, until steady-state is achieved, which may take many time-steps. Here, we describe the implementation of a fully-implicit backward Euler numerical scheme. Higher-order implicit methods are not considered here because the uncertainty associated with boundary conditions and the fitting parameters in the Van Genuchten models (eq. \ref{eq:van-genuchten}) have much more effect than the order of the numerical method used.

The discretized approximation to the mixed-form of Richards equation, using fully-implicit backward Euler, reads:

\begin{equation}
\label{eq:richards-mixed-discrete-time}
    F(\boldsymbol{\psi}^{n+1},\boldsymbol{\psi}^n) = \frac{
    \boldsymbol{\theta}(\boldsymbol{\psi}^{n+1}) - \boldsymbol{\theta}(\boldsymbol{\psi}^n)
    }{\Delta t}
    - \mathbf{D}
        \text{ diag}
        \left(
            \mathbf{k}_{Av}(\boldsymbol{\psi}^{n+1})
        \right)
    \mathbf{G} \boldsymbol{\psi}^{n+1}
    - \mathbf{G}_z
        \left(
            \mathbf{k}_{Av}(\boldsymbol{\psi}^{n+1})
        \right)
    =
    0
\end{equation}
This is a nonlinear system of equations for $\bfpsi\nn$ that needs to be solved numerically by some iterative process. Either a Picard iteration (as in \cite{Celia1990}) or a Newton root-finding iteration with a step length control can be used to solve the system. Note that to deal with dependence of $\theta$ with respect to $\psi$ in Newton's method, we require the computation of $\deriv{\bftheta}{\bfpsi}$. We can complete this computation by using the analytic form of the hydraulic conductivity and water content functions (e.g. derivatives of eq. \ref{eq:van-genuchten}). We note that a similar approach can be used for any smooth curve, even when the connection between $\theta$ and $\psi$ are determined empirically (for example, when $\theta(\psi)$ is given by a spline interpolation of field data).

\subsection{Solving the nonlinear equations}
Regardless of the empirical relation chosen, we must solve \ref{eq:richards-mixed-discrete-time} using an iterative root-finding technique. Newton's method iterates over $m=1,2,\dots$ until a satisfactory estimation of $\bfpsi^{n+1}$ is obtained. Given $\bfpsi\nnm$, we approximate $\FF(\bfpsi^{n+1},\bfpsi\n)$ as:

\begin{equation}
\label{eq:richards-newton}
F(\boldsymbol{\psi}^{n+1},\boldsymbol{\psi}^n)
\approx
F(\boldsymbol{\psi}^{n,m},\boldsymbol{\psi}^n) +
\mathbf{J}_{\boldsymbol{\psi}^{n+1,m}} \delta \boldsymbol{\psi}
\end{equation}
where the Jacobian for iteration, $m$, is:

\begin{equation}
\label{eq:richards-newton-deriv-setup}
\mathbf{J}_{\psi^{n+1,m}}  = \left. {\frac {\partial F(\boldsymbol{\psi},\boldsymbol{\psi}^n)}{\partial \boldsymbol{\psi}}} \right|_{\boldsymbol{\psi}^{n+1,m}}
\end{equation}
The Jacobian is a large dense matrix, and its computation necessitates the computation of the derivatives of $\FF(\bfpsi\nnm,\bfpsi\n)$. We can use numerical differentiation in order to evaluate the Jacobian (or its product with a vector). However, in the context of the inverse problem, an exact expression is preferred. Given the discrete forward problem, we obtain that:

\begin{equation}
\label{eq:richards-newton-deriv}
    \bfJ_{ \bfpsi\nnm} =
    \frac{1}{\Delta t} \deriv{\bftheta(\bfpsi\nnm)}{\bfpsi\nnm}
    -
    \deriv{}{\bfpsi\nnm}
    \left(
    \bfD
        \diag{\bfk_{Av}(\bfpsi\nnm)}
    \mathbf{G}\bfpsi\nnm
    \right)
    -
    \mathbf{G}_z
    \deriv{\bfk_{Av}(\bfpsi\nnm)}{\bfpsi\nnm}
\end{equation}
Here, recall that $\bfk_{Av}$ is harmonically averaged and its derivative can be obtained by the chain rule:

\begin{equation}
\label{eq:mass-matrix-kav-deriv}
\deriv{\mathbf{k}_{Av}(\boldsymbol{\psi})}{\boldsymbol{\psi}}
=
{\rm diag}\left(
    (\mathbf{A}_v \mathbf{k}^{-1}(\boldsymbol{\psi}))^{-2} \right)
\mathbf{A}_v
\diag{
    \mathbf{k}^{-2}(\boldsymbol{\psi}) }
\deriv{\mathbf{k}(\boldsymbol{\psi})}{\boldsymbol{\psi}}
\end{equation}
Similarly, for the second term in (\ref{eq:richards-newton-deriv}) we obtain:

\begin{equation}
\label{eq:richards-newton-deriv-detail}
\frac{\partial}{\partial\boldsymbol{\psi}}
\left(
\mathbf{D}
    \text{ diag}\left(\mathbf{k}_{Av}(\boldsymbol{\psi})\right)
\mathbf{G} \boldsymbol{\psi}
\right)
=
\mathbf{D}
    \text{ diag}\left(\mathbf{k}_{Av}(\boldsymbol{\psi})\right)
\mathbf{G}
+
\mathbf{D}
    \text{ diag}\left(\mathbf{G}\boldsymbol{\psi}\right)
\frac{\partial\mathbf{k}_{Av}(\boldsymbol{\psi})}{\partial\boldsymbol{\psi}}
\end{equation}
Here the notation $\nnm$ has been dropped for brevity. For the computations above, we need the derivatives of functions $\bfk(\bfpsi)$ and $\bftheta(\bfpsi)$; note that, since the relations are assumed local (point wise in space) given the vector, $\bfpsi$, these derivatives are diagonal matrices. For Newton's method, we solve the linear system:

\begin{equation}
\label{eq:richards-newton-update}
\mathbf{J}_{\psi^{n+1,m}}\, \delta \boldsymbol{\psi}\ =\ -  F(\boldsymbol{\psi}^{n+1,m},\boldsymbol{\psi}^n)
\end{equation}

For small-scale problems, we can solve the linear system using direct methods; however, for large-scale problems, iterative methods are more commonly used. The existence of an advection term in the PDE results in a non-symmetric linear system. Thus, when using iterative techniques to solve this system, an appropriate iterative method, such as \textsc{bicgstab} or \textsc{gmres} \citep{templates}, must be used.

At this point, it is interesting to note the difference between the Newton iteration and the Picard iteration suggested in \citep{Celia1990}. We can verify that the Picard iteration uses an approximation to the Jacobian $\bfJ_{\psi\nnm}\, \delta \bfpsi$ given by dropping the second term from \eqref{eq:richards-newton-deriv-detail}. This term can have negative eigenvalues and dropping it is typically done when considering the lagged diffusivity method \citep{vogelBook}. However, as discussed in \citep{vogelBook}, ignoring this term can slow convergence.

Finally, a new iterate is computed by adding the Newton update to the last iterate:

$$\bfpsi\nnmm = \bfpsi\nnm+ \alpha \delta \psi$$
where $\alpha$ is a parameter that guarantees that

$$ \|\FF(\bfpsi^{n,m+1},\bfpsi\n) \| < \| \FF(\bfpsi^{n,m},\bfpsi\n) \| $$
To obtain $\alpha$, we perform an Armijo line search \citep{nw}. In our numerical experiments, we have found that this method can fail when the hydraulic conductivity is strongly discontinuous and changes rapidly. In such cases, Newton's method yields a poor descent direction. Therefore, if the Newton iteration fails to converge to a solution, the update is performed with the mixed-form Picard iteration. Note that Picard iteration can be used, even when Newton’s method fails, because Picard iteration always yields a descent direction \citep{vogelBook}.

At this point, we have discretized Richards equation in both time and space while devoting special attention to the derivatives necessary in Newton's method and the Picard iteration as described in \citep{Celia1990}. The exact derivatives of the discrete problem will be used in the following two sections, which outline the implicit formula for the sensitivity and its incorporation into a general inversion algorithm.
\section{Inverse Problem}
\label{sec:richards-inverse}

The location and spatial variability of, for example, an infiltration front over time is inherently dependent on the hydraulic properties of the soil column. As such, direct or proxy measurements of the water content or pressure head at various depths along a soil profile contain information about the soil properties. We pose the inverse problem, which is the estimation of distributed hydraulic parameters, given either water content or pressure data. We frame this problem under the assumption that we wish to estimate hundreds of thousands to millions of distributed model parameters. Due to the large number of model parameters that we aim to estimate in this inverse problem, Bayesian techniques or external numerical differentiation, such as the popular PEST toolbox \citep{Doherty2004}, are not computationally feasible. Instead, we will employ a direct method by calculating the exact derivatives of the discrete Richards equation and solving the sensitivity implicitly. For brevity, we show the derivation of the sensitivity for an inversion model of only saturated hydraulic conductivity, $K_s$, from pressure head data, $\bfdo$. This derivation can be readily extended to include the use of water content data and inverted for other distributed parameters in the heterogeneous hydraulic conductivity function. We will demonstrate the sensitivity calculation for multiple distributed parameters in the  numerical examples (Section~\ref{sec:richards-examples}).

The Richards equation simulation produces a pressure head field at all points in space as well as through time. Data can be predicted, $\bfdp$, from these fields and compared to observed data, $\bfdo$. To be more specific, we let $\bfPsi = [(\bfpsi^{1})^{\top},\ldots,(\bfpsi^{n_{t}})^{\top}]^{\top}$ be the (discrete) pressure field for all space and $n_{t}$ time steps. When measuring pressure head recorded only in specific locations and times, we define the predicted data, $\bfdp$, as $\bfdp = \bfP \bfPsi(\bfm)$. Here, the vector $\mathbf{m}$ is the vector containing all of the parameters which we are inverting for (e.g. $K_s, \alpha, n, \theta_r$, or $\theta_s$ when using the van Genuchten empirical relation). The matrix, $\bfP$, interpolates the pressure head field, $\bfPsi$, to the locations and times of the measurements. Since we are using a simple finite volume approach and backward Euler in time, we use linear interpolation in both space and time to compute $\bfdp$ from $\bfPsi$. Thus, the entries of the matrix $\bfP$ contain the interpolation weights. For linear interpolation in 3D, $\bfP$ is a sparse matrix that contains up to eight non-zero entries in each row. Note that the time and location of the data measurement is independent and decoupled from the numerical discretization used in the forward problem. A water retention curve, such as the van Genuchten model, can be used for computation of predicted water content data, which requires another nonlinear transformation,  $\bfdp^\theta = \bfP \bftheta(\bfPsi(\bfm), \bfm)$. Note here that the transformation to water content data, in general, depends on the model to be estimated in the inversion, which will be addressed in the numerical examples. For brevity in the derivation that follows, we will make two simplifying assumptions: (1) that the data are pressure head measurements, which requires a linear interpolation that is not dependent on the model; and, (2) that the model vector, $\bfm$, describes only distributed saturated hydraulic conductivity. Our software implementation \emph{does not} make these assumptions; our numerical examples will use water content data, a variety of empirical relations, and calculate the sensitivity to multiple heterogeneous empirical parameters.

\subsection{Solution through optimization}
We can now formulate the discrete inverse problem to estimate saturated hydraulic conductivity, $\bfm$, from the observed pressure head data, $\bfdo$. We frame the inversion as an optimization problem, which minimizes a data misfit and a regularization term.

\begin{equation}
\label{eq:richards-objective}
\widehat{\mathbf{m}} =
{\rm arg}\min_{\mathbf{m}}\ \Phi ( \mathbf{m} ) =
\frac{1}{2} \left\|\mathbf{W}_d (\mathbf{d}_{\rm pred}(\mathbf{m})- \mathbf{d}_{\rm obs}) \right\|^2_2+
\frac { \beta  }{ 2 } { \left\| \mathbf{W}_m(\mathbf{m}- \mathbf{m} _{\rm ref }) \right\|  }^2_2.
\end{equation}
The first term in the objective function is the data misfit; it contains a weighted difference between the observed and predicted data. Assuming that the observed data is noisy, with independent distributed Gaussian noise having standard deviation $\boldsymbol{\sigma}$, the weighting matrix, $\bfW_d$, is a diagonal matrix that contains the entries $ \boldsymbol{\sigma}_{i}^{-1}$ on its diagonal. The matrix, $\bfW_m$, is a regularization matrix that is chosen based on a-priori information and assumptions about the geologic setting \citep{DougTutorial}. In a regularized inversion framework, if we estimate only the log of the static conductivity, $K_{s}$, in \eqref{eq:van-genuchten}, then we typically choose $\bfW_m$  as a combination of the first order gradient matrix and an identity matrix. In a Bayesian framework, where one assumes that $\bfm$ is Gaussian, the matrix $\bfW_m^\top \bfW_m$ is the inverse covariance matrix. If $\bfm$ contains more than a single distributed hydrogeologic parameter, then a correlation term is typically added \citep{HaberHoltzman2013}. The reference model, $\bfm_{\rm ref}$, is chosen based on a-priori information about the background, $\bfm$. Finally, we add the misfit and regularization terms with a regularization parameter, $\beta$, which balances the a-priori information, or smoothness, with the fit to noisy data. This is the standard approach in geophysical inversions \cite{tikhonov1977, DougTutorial, Constable1987, haber2015computational} where hundreds of thousands to millions of distributed parameters are commonly estimated in a deterministic inversion. The hydrogeologic literature also shows the use of these techniques; however, there is also a large community advancing stochastic inversion techniques and geologic realism (cf. \cite{Linde2015}). Regardless of the inversion algorithm used, an efficient method to calculate the sensitivity is crucial; this method is the focus of our work.

To minimize the objective function, we use an inexact Gauss-Newton method that is well-suited for large-scale problems \citep{nw}. For the Gauss-Newton method, the gradient of Equation~\ref{eq:richards-objective} is computed and iteratively driven towards $0$ by using an approximate Hessian, $\bfH$. For the problem at hand, the gradient can be expressed as:

\begin{equation}
\label{eq:richards-objective-deriv}
\nabla_{\mathbf{m}} \Phi ( \mathbf{m} ) =  \mathbf{J}^\top \mathbf{W}_d^\top \mathbf{W}_d(\mathbf{d}_{\rm pred}(\mathbf{m}) - \mathbf{d}_{\rm obs})+\beta \mathbf{W}_m^\top \mathbf{W}_m(\mathbf{m}-\mathbf{m}_{\rm ref})= 0
\end{equation}
where the sensitivity matrix, $\bfJ = \nabla_{\bfm} \bfdp(\bfm)$, is the derivative of the predicted data with respect to the model parameters. The Hessian of the objective function, $\bfH$, is approximated with the first order terms and is guaranteed to be positive definite.

\begin{equation}
\label{eq:richards-objective-deriv2}
\nabla^2_{\mathbf{m}} \Phi ( \mathbf{m} )
\approx \mathbf{H} =
\mathbf{J}^\top \mathbf{W}_d^\top \mathbf{W}_d \mathbf{J} +
\beta \mathbf{W}_m^\top \mathbf{W}_m = 0
\end{equation}
Finally, the (inexact) Gauss-Newton update, $\delta \bfm$, is computed by (approximately) solving the system

$$\bfH \delta \bfm = -\nabla_{\bfm} \Phi ( \bfm )$$

using some iterative technique. In this work, we use the preconditioned conjugate gradient (PCG) method, allowing us to work with large-scale problems. To precondition the conjugate gradient algorithm, we use a standard limited-memory BFGS algorithm initiated with the inverse of $\bfW_m^\top\bfW_m$ instead of the identity. For more details on the preconditioner, see \citep{haber2}.

It is important to note that the sensitivity matrix, $\bfJ$, as well as the approximate Hessian, $\bfH$, are large, dense matrices and their explicit computations result in an algorithm that is constrained by computational memory. However, as we show in the following section, one can avoid this computation and efficiently compute the products of $\bfJ$ and $\bfJ^{\top}$ with a vector. This computation, coupled with an iterative optimization method, results in an efficient algorithm.

Once the model update, $\delta \bf m$, is found, a Gauss-Newton step is taken:

$$\bfm_{k+1} = \bfm_{k} + \alpha_k \delta \bfm$$

where $\alpha_k$ is the current line search parameter. We use a backtracking Armijo line search \citep{Armijo1966} to enforce sufficient decrease in our objective function \citep{nw}. We repeat the minimization procedure, which updates the linearization until the norm of the gradient falls below a certain tolerance, the model update is sufficiently small, or until we have exhausted the maximum number of iterations. Values for these stopping criteria are often problem-dependent. Running synthetic models with elements similar to observed data can often give insight into appropriate stopping criteria \citep{Aster2004}.


\subsection{Implicit sensitivity calculation}
\label{sec:richards-derivs}

The optimization problem requires the derivative of the pressure head with respect to the model parameters, $\frac{\partial\bfPsi}{\partial\bfm}$. We can obtain an approximation of the sensitivity matrix  through a finite difference method on the forward problem \citep{Simunek1996, Finsterle2011, Finsterle2011c}. One forward problem, or two, when using central differences, must be completed for each column in the Jacobian at every iteration of the optimization algorithm. This style of differentiation proves advantageous in that it can be applied to any forward problem; however, it is highly inefficient and introduces errors into the inversion that may slow the convergence of the scheme \citep{Doherty2004}. Automatic differentiation (AD) can also be used \citep{nw}. However, AD does not take the structure of the problem into consideration and requires that the dense Jacobian be explicitly formed. \cite{Bitterlich2002} presents three algorithms (finite difference, adjoint, and direct) to directly compute the elements of the dense sensitivity matrix for Richards equation. As problem size increases, the memory required to store this dense matrix often becomes a practical computational limitation \citep{hao2, Towara2015}. As we show next, it is possible to explicitly write the derivatives of the Jacobian and evaluate their products with vectors using only sparse matrix operations. This algorithm is much more efficient than finite differencing or AD, especially for large scale simulations, since it does not require explicitly forming and storing a large dense matrix. Rather, the algorithm efficiently computes matrix-vector and adjoint matrix-vector products with sensitivity. We can use these products for the solution of the Gauss-Newton system when using the conjugate gradient method, which bypasses the need for the direct calculation of the sensitivity matrix and makes solving large-scale inverse problems possible. Other geophysical inverse problems have used this idea extensively, especially in large-scale electromagnetics (cf. \cite{hao}). The challenge in both the derivation and implementation for Richards equation lies in differentiating the nonlinear time-dependent forward simulation with respect to multiple distributed hydraulic parameters.

The approach to implicitly constructing the derivative of Richards equation in time involves writing the whole time-stepping process as a block bi-diagonal matrix system. The discrete Richards equation can be written as a function of the model. For a single time-step, the equation is written:

\begin{align}
\label{eq:richards-timestep}
F(\boldsymbol{\psi}^{n+1}(\mathbf{m}),\boldsymbol{\psi}^n(\mathbf{m}),\mathbf{m}) =
\frac{\boldsymbol{\theta}^{n+1}(\boldsymbol{\psi}^{n+1}) - \boldsymbol{\theta}^n(\boldsymbol{\psi}\n)}{\Delta t}
\nonumber\\-
\ \mathbf{D}
    \text{ diag}\left(\mathbf{k}_{Av}(\boldsymbol{\psi}^{n+1},\mathbf{m})\right)
    \mathbf{G} \boldsymbol{\psi}^{n+1}
-
\mathbf{G}_{z} \mathbf{k}_{Av}(\boldsymbol{\psi}^{n+1},\mathbf{m})
= 0
\end{align}
In this case, $\bfm$ is a vector that contains all the parameters of interest. Note that $\bfpsi\nn$ and $\bfpsi\n$ are also functions of $\bfm$. In general, $\bftheta\nn$ and $\bftheta\n$ are also dependent on the model; however, for brevity, we will omit these derivatives. The derivatives of $\FF$ to the change in the parameters $\bfm$ can be written as:

\begin{equation}
\label{eq:richards-timestep-chain}
    \nabla_\mathbf{m}  F(\boldsymbol{\psi}^n,\boldsymbol{\psi}^{n+1},\mathbf{m})
    =
    \frac{\partial F}{\partial \mathbf{k}_{Av}}
    \frac{\partial\mathbf{k}_{Av}}{\partial\mathbf{m}}
    + \frac{\partial F}{\partial \boldsymbol{\psi}^n}\frac{\partial\boldsymbol{\psi}^n}{\partial\mathbf{m}}
    + \frac{\partial F}{\partial \boldsymbol{\psi}^{n+1}}\frac{\partial\boldsymbol{\psi}^{n+1}}{\partial\mathbf{m}}
    =0
\end{equation}
or, in more detail:

\begin{align}
\label{eq:richards-timestep-deriv}
\frac{1}{\Delta t}
\left(
    \frac{\partial \boldsymbol{\theta}^{n+1}}{\partial\boldsymbol{\psi}^{n+1}}
    \frac{\partial \boldsymbol{\psi}^{n+1}}{\partial\mathbf{m}}
    -
    \frac{\partial \boldsymbol{\theta}^n}{\partial\boldsymbol{\psi}^n}
    \frac{\partial \boldsymbol{\psi}^n}{\partial\mathbf{m}}
\right)
-
\mathbf{D}
    \text{ diag}\left( \mathbf{G} \boldsymbol{\psi}^{n+1} \right)
    \left(
        \frac{\partial \mathbf{k}_{Av}}{\partial\mathbf{m}} +
        \frac{\partial \mathbf{k}_{Av}}{\partial\boldsymbol{\psi}^{n+1}}
        \frac{\partial \boldsymbol{\psi}^{n+1}}{\partial\mathbf{m}}
    \right)
\nonumber\\
-\
\mathbf{D}
    \text{ diag}\left( \mathbf{k}_{Av}(\boldsymbol{\psi}^{n+1}) \right)
\mathbf{G}
\frac{\partial \boldsymbol{\psi}^{n+1}}{\partial\mathbf{m}}
-
\mathbf{G}_{z}
\left(
    \frac{\partial \mathbf{k}_{Av}}{\partial\mathbf{m}} +
    \frac{\partial \mathbf{k}_{Av}}{\partial\boldsymbol{\psi}^{n+1}}
    \frac{\partial \boldsymbol{\psi}^{n+1}}{\partial\mathbf{m}}
\right)
& = 0
\end{align}
The above equation is a linear system of equations and, to solve for $\deriv{\bfPsi}{\bfm}$, we rearrange the block-matrix equation:

\begin{align}
\label{eq:richards-timestep-deriv-blocks}
\overbrace{
    \left[
        \frac{1}{\Delta t}
        \frac{\partial \boldsymbol{\theta}^{n+1}}{\partial\boldsymbol{\psi}^{n+1}}
        -\mathbf{D}
        \text{ diag}\left( \mathbf{G} \boldsymbol{\psi}^{n+1} \right)
        \frac{\partial \mathbf{k}_{Av}}{\partial\boldsymbol{\psi}^{n+1}}
        -\mathbf{D}
        \text{ diag}\left( \mathbf{k}_{Av}(\boldsymbol{\psi}^{n+1},\mathbf{m}) \right)
        \mathbf{G}
        - \mathbf{G}_{z}
        \frac{\partial \mathbf{k}_{Av}}{\partial\boldsymbol{\psi}^{n+1}}
    \right]
}^{\mathbf{A}_0(\boldsymbol{\psi}^{n+1})}
\frac{\partial \boldsymbol{\psi}^{n+1}}{\partial\mathbf{m}}
\nonumber\\
+
\underbrace{
    \left[
        -\frac{1}{\Delta t}
        \frac{\partial \boldsymbol{\theta}^n}{\partial\boldsymbol{\psi}^n}
    \right]
}_{\mathbf{A}_{-1}(\boldsymbol{\psi}^n)}
\frac{\partial \boldsymbol{\psi}^n}{\partial\mathbf{m}}
=
\underbrace{
\left[
    -\mathbf{D}
    \text{ diag}\left( \mathbf{G} \boldsymbol{\psi}^{n+1} \right)
    \frac{\partial \mathbf{k}_{Av}}{\partial\mathbf{m}}
    -\mathbf{G}_{z}
    \frac{\partial \mathbf{k}_{Av}}{\partial\mathbf{m}}
\right]
}_{\mathbf{B}(\psi^{n+1})}&
\end{align}
Here, we use the subscript notation of $\bfA_0(\bfpsi\nn)$ and $\bfA_{-1}(\bfpsi\n)$ to represent two block-diagonals of the large sparse matrix $\bfA({\bfPsi},\bfm)$. Note that all of the terms in these matrices are already evaluated when computing the Jacobian of Richards equations in Section~\ref{sec:richards-forward} and that they contain only basic sparse linear algebra manipulations without the inversion of any matrix. If $\bfpsi_0$ does not depend on the model, meaning the initial conditions are independent, then we can formulate the block system as:

\begin{equation}
\label{eq:richards-timestep-deriv-matrix}
\overbrace{
    \left[
    \begin{array}{cccc}
        \mathbf{A}_0(\boldsymbol{\psi}_1)\\
        \mathbf{A}_{-1}(\boldsymbol{\psi}_1)&\mathbf{A}_0(\boldsymbol{\psi}_2)\\
        &\mathbf{A}_{-1}(\boldsymbol{\psi}_2)&\mathbf{A}_0(\boldsymbol{\psi}_3)\\
        &\qquad\ddots&\qquad\ddots\\
        &&\mathbf{A}_{-1}(\boldsymbol{\psi}_{n_{t}-1})&\mathbf{A}_0(\boldsymbol{\psi}_{n_{t}})\\
    \end{array}
    \right]
}^{\mathbf{A}(\boldsymbol{\Psi},\mathbf{m})}
\overbrace{
    \left[
    \begin{array}{c}
        \frac{\partial \boldsymbol{\psi}_1}{\partial \mathbf{m}}\\
        \frac{\partial \boldsymbol{\psi}_2}{\partial \mathbf{m}}\\
        \vdots\\
        \frac{\partial \boldsymbol{\psi}_{n_{t}-1}}{\partial \mathbf{m}}\\
        \frac{\partial \boldsymbol{\psi}_{n_{t}}}{\partial \mathbf{m}}\\
    \end{array}
    \right]
}^{\frac{\partial\boldsymbol{\Psi}}{\partial\mathbf{m}}}
    =
\overbrace{
    \left[
    \begin{array}{c}
        \mathbf{B_1}(\boldsymbol{\psi}_1)\\
        \mathbf{B_2}(\boldsymbol{\psi}_2)\\
        \vdots\\
        \mathbf{B_{n-1}}(\boldsymbol{\psi}_{n_{t}-1})\\
        \mathbf{B_n}(\boldsymbol{\psi}_{n_{t}})\\
    \end{array}
    \right]
}^{\mathbf{B}(\boldsymbol{\Psi},\mathbf{m})}
\end{equation}
This is a block matrix equation; both the storage and solve will be expensive if it is explicitly computed. Indeed, its direct computation is equivalent to the adjoint method \citep{Bitterlich2002, DeanChen2011}.

Since only matrix vector products are needed for the inexact Gauss-Newton optimization method, the  matrix $\bfJ$ is never needed explicitly  and only the products of the form $\bfJ \bfv$ and $\bfJ^\top \bfz$ are needed for arbitrary vectors $\bfv$ and $\bfz$. Projecting the full sensitivity matrix onto the data-space using $\bfP$ results in the following equations for the Jacobian:

\begin{subequations}
\begin{equation}
\label{eq:richards-timestep-deriv-mult}
    \mathbf{J} = \mathbf{P} \mathbf{A}(\boldsymbol{\Psi},\mathbf{m})^{-1} \mathbf{B}(\boldsymbol{\Psi},\mathbf{m})
\end{equation}
\begin{equation}
\label{eq:richards-timestep-deriv-mult-adjoint}
    \mathbf{J}^\top =   \mathbf{B}(\boldsymbol{\Psi},\mathbf{m})^\top \mathbf{A}(\boldsymbol{\Psi},\mathbf{m})^{-\top} \mathbf{P}^\top
\end{equation}
\end{subequations}
In these equations, we are careful to not write $\deriv{\bfPsi}{\bfm}$, as it is a large dense matrix which we do not want to explicitly compute or store. Additionally, the matrices $\bf A(\bfPsi,\bfm)$ and $\bf B(\bfPsi,\bfm)$ do not even need to be explicitly formed because the matrix $\bf A(\bfPsi,\bfm)$ is a triangular block-system, which we can solve using forward or backward substitution with only one block-row being solved at a time (this is equivalent to a single time step). To compute the matrix vector product, $\bfJ \bfv$, we use a simple algorithm:

\begin{enumerate}
    \item Given the vector $\bfv$ calculate $\bfy = \mathbf{Bv}$
    \item Solve the linear system $\mathbf{Aw} = \bfy$ for the vector $\bfw$
    \item Set $\bfJ \bfv = \bfP \bfw$
\end{enumerate}

Here, we note that we complete steps (1) and (2) using a for-loop with only one block-row being computed and stored at a time. As such, only the full solution, $\bfPsi$, is stored and all other block-entries may be computed as needed. There is a complication here if data is in terms of water content or effective saturation, as the data projection is no longer linear and may have model dependence. These complications can be dealt with using the chain rule during step (3). Similarly, to compute the adjoint $\bfJ^\top \bfz$ involves the intermediate solve for $\bfy$ in $\bfA^\top \bfy = \bfP^\top \bfz$ and then computation of $\bfB^\top \bfy$. Again, we solve the block-matrix via backward substitution with all block matrix entries being computed as needed. Note that the backward substitution algorithm can be viewed as time stepping, which means that it moves from the final time back to the initial time. This time stepping is equivalent to the adjoint method that is discussed in \cite{DeanChen2011} and references within. The main difference between our approach and the classical adjoint-based method is that our approach yields the {\em exact} gradient of the discrete system; no such guarantee is given for adjoint-based methods.

The above algorithm and the computations of all of the different derivatives summarizes the technical details of the computations of the sensitivities. Equipped with this ``machinery'', we now demonstrate that validity of our implementation and our ability to solve large-scale problems using modest computational resources.

\section{Validation}
\label{sec:richards-validation}

\subsection{Forward problem}
\begin{figure}[!ht]
\begin{center}
\includegraphics[width=0.48\textwidth]{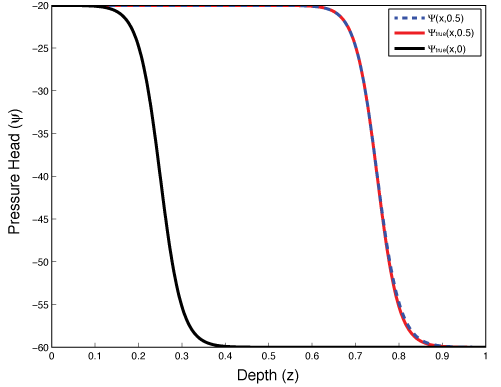}
\end{center}
\caption{Fictitious source test in 1D showing the analytic function $\Psi_{\text{true}}$ at times 0.0 and 0.5 and the numerical solution $\Psi(x,0.5)$ using the mixed-form Newton iteration.}
\label{fig:richards-validation-source}
\end{figure}

Richards equation has no analytic solution, which makes testing the code more involved. Code-to-code comparisons have been completed for comparison to \cite{Celia1990}, which can be found in \cite{richardsceliacomparison}. Here we have chosen to use a fictitious source experiment to rigorously test the code. In this experiment, we approximate an infiltration front by an arctangent function in one dimension, which is centered over the highly nonlinear part of the van Genuchten curves, with $\psi\in[-60,-20]$ centimeters. The arctangent curve advects into the soil column with time. The advantage of using an analytic function is that the derivative is known explicitly and can be calculated at all times. However, it should be noted that Richards equation does not satisfy the analytic solution exactly, but differs by a function, $S(x,t)$. We refer to this function as the fictitious source term. The analytic function that we used has similar boundary conditions and shape to an example in \cite{Celia1990} and is considered over the domain $x\in[0, 1]$.
\begin{equation}
\label{eq:fictitiousSource}
    \Psi_{\text{true}}(x,t)=-20\arctan(20((x-0.25)-t))-40
\end{equation}
This analytic function is shown at times 0 and 0.5 in Figure \ref{fig:richards-validation-source} and has a pressure head range of $\psi\in[-60,-20]$. We can compare these values to the van Genuchten curves in Figure \ref{fig:van-genuchten}. We can then put the known pressure head into Richards equation (\ref{eq:richards-mixed}) and calculate the analytic derivatives and equate them to a source term, $S(x,t)$. Knowing this source term and the analytic boundary conditions, we can solve discretized form of Richards equation, which should reproduce the analytic function in Equation \ref{eq:fictitiousSource}. Table \ref{table:richards-source} shows the results of the fictitious source test when the number of mesh-cells is doubled and the time-discretization is both fixed and equivalent to the mesh size (i.e. $k=h$). In this case, we expect that the backward-Euler time discretization, which is $\mathcal{O}(\delta)$, will limit the order of accuracy. The final column of Table \ref{table:richards-source} indeed shows that the order of accuracy is $\mathcal{O}(\delta)$. The higher errors in the coarse discretization are due to high discontinuities and changes in the source term, which the coarse discretization does not resolve.

\begin{table}[!ht]
\centering
\caption{Fictitious source test for Richards equation in 1D using the mixed-form Newton iteration.}
\begin{tabular}{*{3}{c}}
\\[-0.5em]
\hline
\\[-0.5em]
Mesh Size (n)    &     $||\Psi(x,0.5) - \Psi_{\text{true}}(x,0.5)||_\infty$     &    Order Decrease, $\mathcal{O}(\delta)$ \\[0.5em]
\hline
\\[-0.5em]
           64    &    5.485569e+00    &                    \\
          128    &    2.952912e+00    &             0.894  \\
          256    &    1.556827e+00    &             0.924  \\
          512    &    8.035072e-01    &             0.954  \\
         1024    &    4.086729e-01    &             0.975  \\
         2048    &    2.060448e-01    &             0.988  \\
         4096    &    1.034566e-01    &             0.994  \\
         8192    &    5.184507e-02    &             0.997  \\

\\[-0.5em]
\hline
\end{tabular}
\label{table:richards-source}
\end{table}

\subsection{Inverse problem}
In order to test the implicit sensitivity calculation, we employ derivative and adjoint tests as described in \cite{haber2015computational}. Given that the Taylor expansion of a function $f(\mathbf{m} + h \Delta \mathbf{m})$ is
\begin{equation}
    f(\mathbf{m}+ h  \Delta \mathbf{m}) = f(\mathbf{m}) + h  \mathbf{J} \Delta \mathbf{m} + \mathcal{O}(h^2),
\end{equation}
for any of the model parameters considered, we see that our approximation of $f(\mathbf{m}+ h  \Delta \mathbf{m})$ by $ f(\mathbf{m}) + h  \mathbf{J} \Delta \mathbf{m}$ should converge as $\mathcal{O}(h^2)$ as $h$ is reduced. This allows us to verify our calculation of $\mathbf{J}\mathbf{v}$. To verify the adjoint, $\mathbf{J}^\top\mathbf{v}$, we check that
\begin{equation}
    \mathbf{w}^\top\mathbf{J}\mathbf{v} = \mathbf{v}^\top \mathbf{J}^\top \mathbf{w}
\end{equation}
for any two random vectors, $\mathbf{w}$ and $\mathbf{v}$. These tests are run for all of the parameters considered in an inversion of Richards equation. Within our implementation, both the derivative and adjoint tests are included as unit tests which are run on any updates to the implementation (\url{https://travis-ci.org/simpeg/simpeg}).

\section{Comparison of iterative methods}
\label{sec:richards-iterative}

\begin{figure}[ht]
\begin{center}
\includegraphics[width=0.9\textwidth]{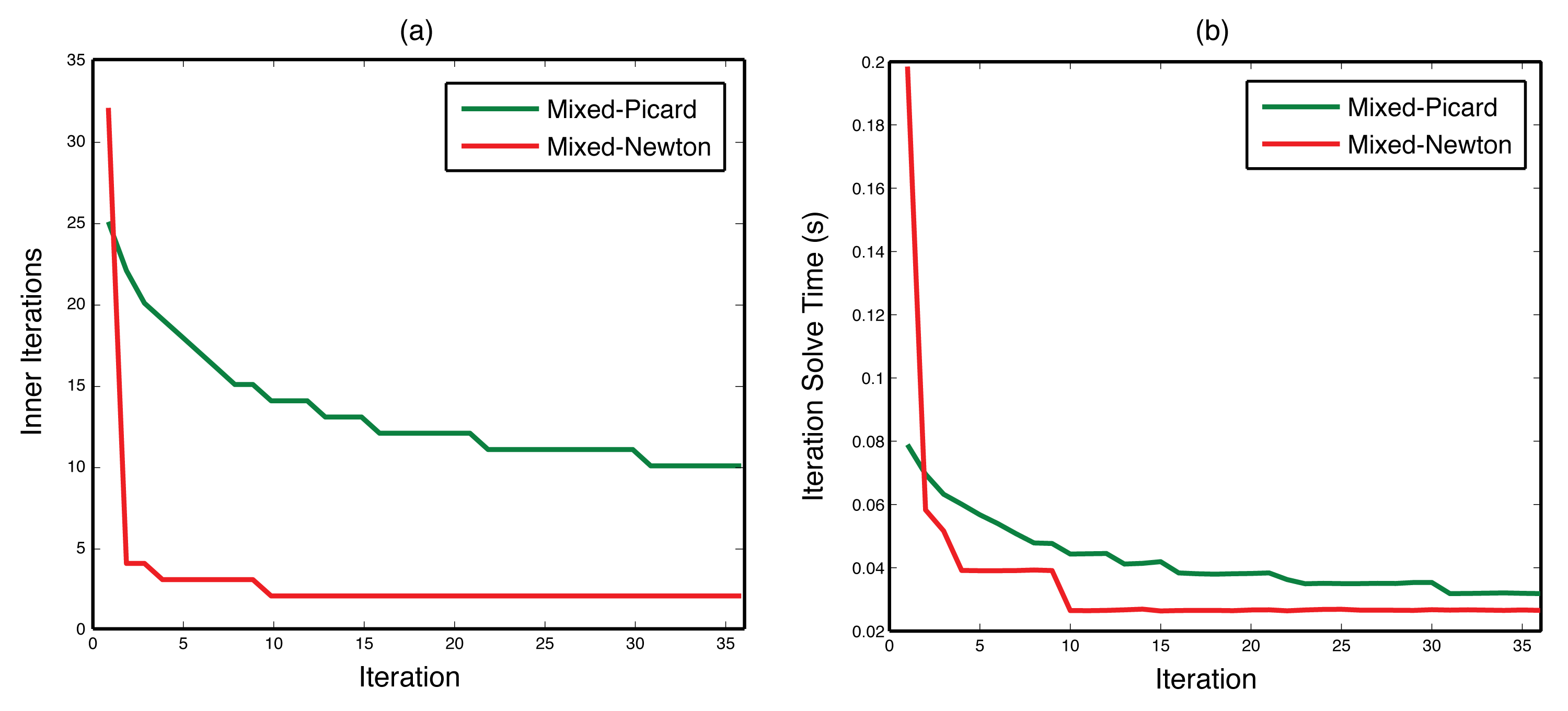}
\end{center}
\caption{Comparison of iterative schemes ($\Delta t = 10$s) showing (a) the number of inner iterations per iteration and (b) the solve time per iteration.}
\label{fig:richards-iterations}
\end{figure}

With the goal of extending these simple one-dimensional examples into three  dimensions, the efficiency of the iterative method used is of critical importance for computational solve time. Figure~\ref{fig:richards-iterations} shows a comparison of the iterative methods used for the numerical example described above (where $\Delta t = 10$s) . Here, both of the Picard schemes take between 10 to 25 iterations per time-step, while the Newton scheme generally uses less than five iterations (Figure~\ref{fig:richards-iterations}a). The difference between iteration counts occurs due to the use of the gradient information in the Newton method, which speeds convergence when the initial guess is close to the solution. When the initial guess is not close to the solution, for example at $t=0$s, the Newton method is slower to converge than the Picard method. The first iteration of Newton has both a higher number of iterations and solve time (Figure~\ref{fig:richards-iterations}b). In both Figure~\ref{fig:richards-iterations}a \& \ref{fig:richards-iterations}b, there is an overall decrease in the number of inner iterations and the time taken per iteration as the system reaches steady-state. The iteration count has previously been used to inform an adaptive time-stepping scheme \citep{RichardsFOAM}. Table~\ref{table:richards-iterations} shows the overall results of solve time and total iterations. The Newton method is marginally faster than the Picard method because it takes the fewest number of iterations.

\begin{table}[!ht]
\centering
\caption{Comparison of total solve time and iteration of numerical schemes ($\Delta t = 10$s)}
\begin{tabular}{*{3}{c}}
\hline
Method Name &  Solve Time (s) &    Total Iterations  \\
\hline
Mixed-Picard   &     1.574    &                 479  \\
Mixed-Newton   &     1.381    &                 112  \\
\hline
\end{tabular}
\label{table:richards-iterations}
\end{table}
\section{Three dimensional inversion}
\label{sec:richards-examples}

In this section, we turn our attention to recovering a three-dimensional soil structure given water content data. The example, motivated by a field experiment introduced in \citep{Pidlisecky2013}, shows a time-lapse electrical resistivity tomography survey completed in the base of a managed aquifer recharge pond. The goal of this management practice is to infiltrate water into the subsurface for storage and subsequent recovery. Such projects require input from geology, hydrology, and geophysics to map the hydrostratigraphy, to collect and interpret time-lapse geophysical measurements, and to integrate all results to make predictions and decisions about fluid movement at the site. As such, the hydraulic properties of the aquifer are important to characterize, and information from hydrogeophysical investigations has been demonstrated to inform management practices \citep{Pidlisecky2013}. We use this context to motivate both the model domain setup of the following synthetic experiment and the subsequent inversion.

\begin{figure}[!htbp]
\begin{center}
\includegraphics[width=0.8\textwidth]{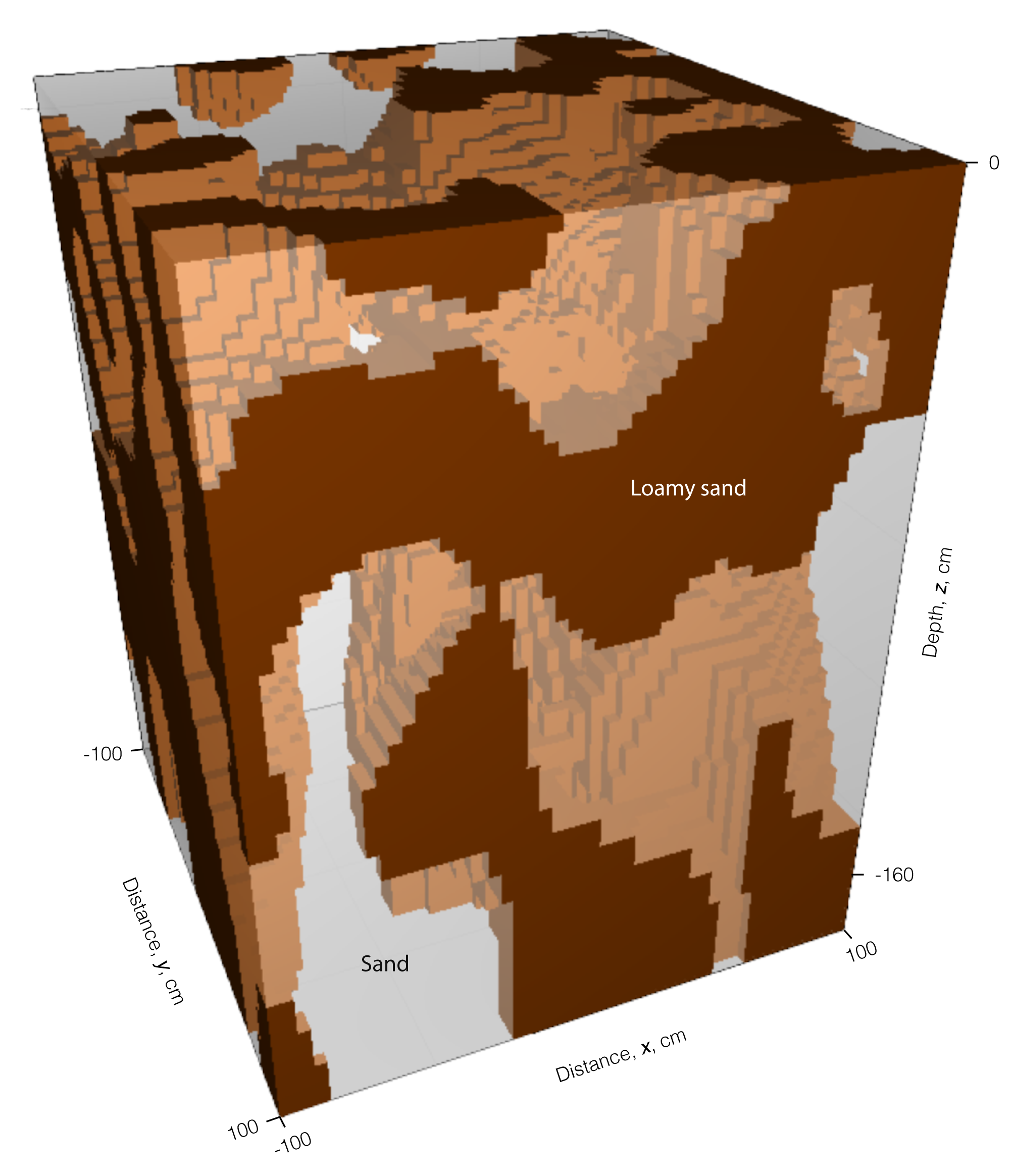}
\end{center}
\caption{
Soil structure in three dimensions showing the boundary between two soil types of sand and loamy sand.
}
\label{fig:richards-3d_model}
\end{figure}

For the inverse problem solved here, we assume that time-lapse water-content information is available at many locations in the subsurface. In reality, water content information may only be available through proxy techniques, such as electrical resistivity methods. These proxy data can be related to hydrogeologic parameters using inversion techniques based solely on the geophysical inputs (cf. \cite{Mawer2013}). For the following numerical experiments, we do not address complications in empirical transformations, such as Archie's equation \citep{Archie1942}. The synthetic numerical model has a domain with dimensions 2.0 m $\times$ 2.0 m $\times$ 1.7 m for the $x$, $y$, and $z$ dimensions, respectively. The finest discretization used is 4 cm in each direction. We use padding cells to extend the domain of the model (to reduce the effect of boundary conditions in the modelling results). These padding cells extend at a factor of 1.1 in the negative $z$ direction. We use an exponentially expanding time discretization with 40 time steps and a total time of 12.3 hours. This choice in discretization leads to a mesh with 1.125$\times10^5$ cells in space ($50 \times 50 \times 45$). To create a three-dimensionally varying soil structure, we construct a model for this domain using a three-dimensional, uniformly random field, $\in [0, 2]$, that is convolved with an anisotropic smoothing kernel for a number of iterations. We create a binary distribution from this random field by splitting the values above and below unity. Figure~\ref{fig:richards-3d_model} shows the resulting model, which reveals potential flow paths. We then map van Genuchten parameters to this synthetic model as either a sand or a loamy-sand. The van Genuchten parameters for sand are: $K_s$: 5.83e-05m/s, $\alpha$: 13.8, $\theta_r$: 0.02, $\theta_s$: 0.417, and  $n$: 1.592; and for loamy-sand are: $K_s$: 1.69e-05m/s, $\alpha$: 11.5, $\theta_r$: 0.035, $\theta_s$: 0.401, and  $n$: 1.474. For this inversion, we are interested in characterizing the soil in three dimensions.

\begin{figure}[!htbp]
\begin{center}
\includegraphics[width=0.9\textwidth]{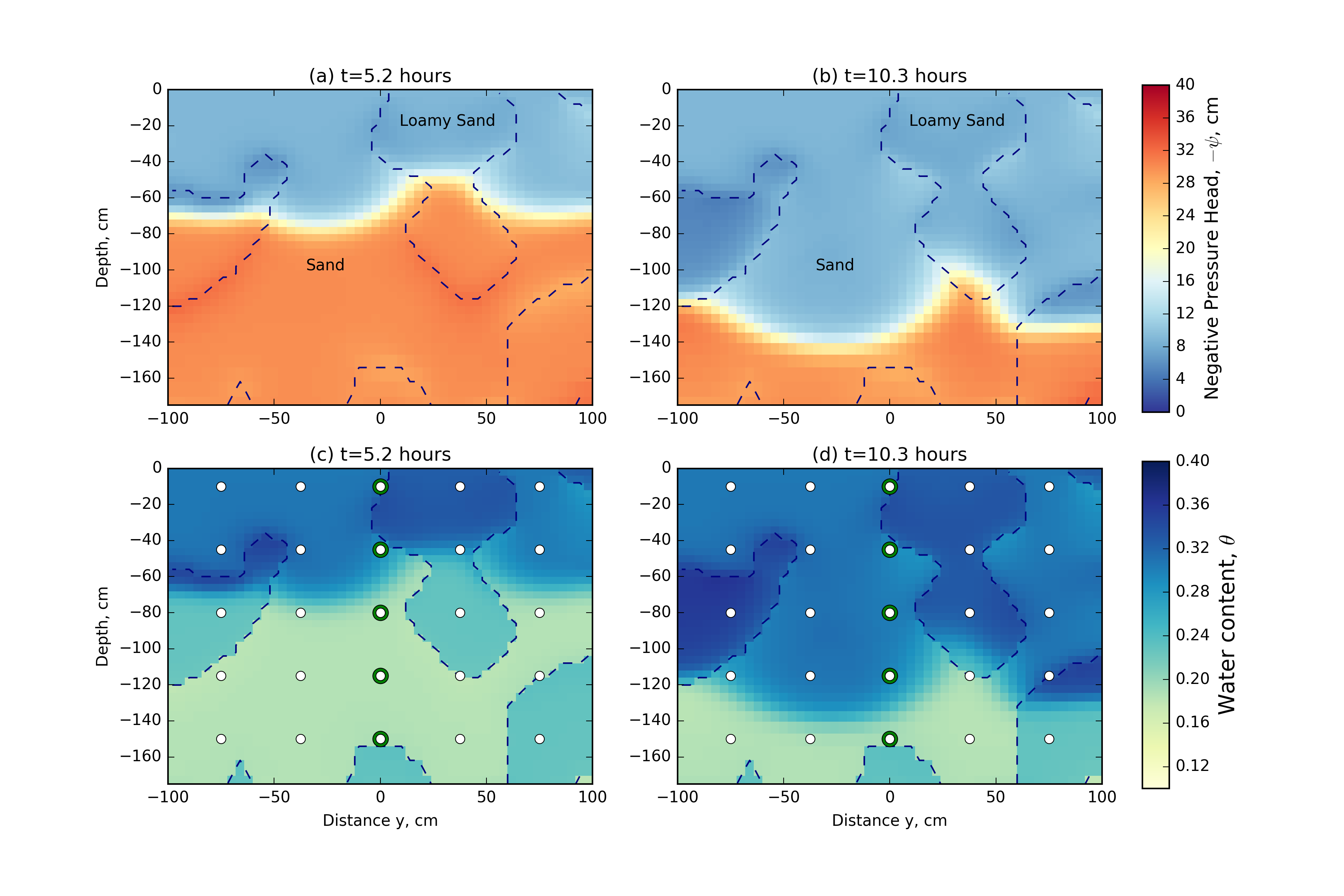}
\end{center}
\caption{Vertical cross sections through the pressure head and saturation fields from the numerical simulation at two times: (a) pressure head field at $t=5.2$ hours and (b) $t=10.3$ hours; and (c) saturation field at $t=5.2$ hours and (d) $t=10.3$ hours. The saturation field plots also show measurement locations and green highlighted regions that are shown in Figure~\ref{fig:richards-data}. The true location of the two soils used are shown with a dashed outline.}
\label{fig:richards-fields}
\end{figure}

\begin{figure}[!htbp]
\begin{center}
\includegraphics[width=0.9\textwidth]{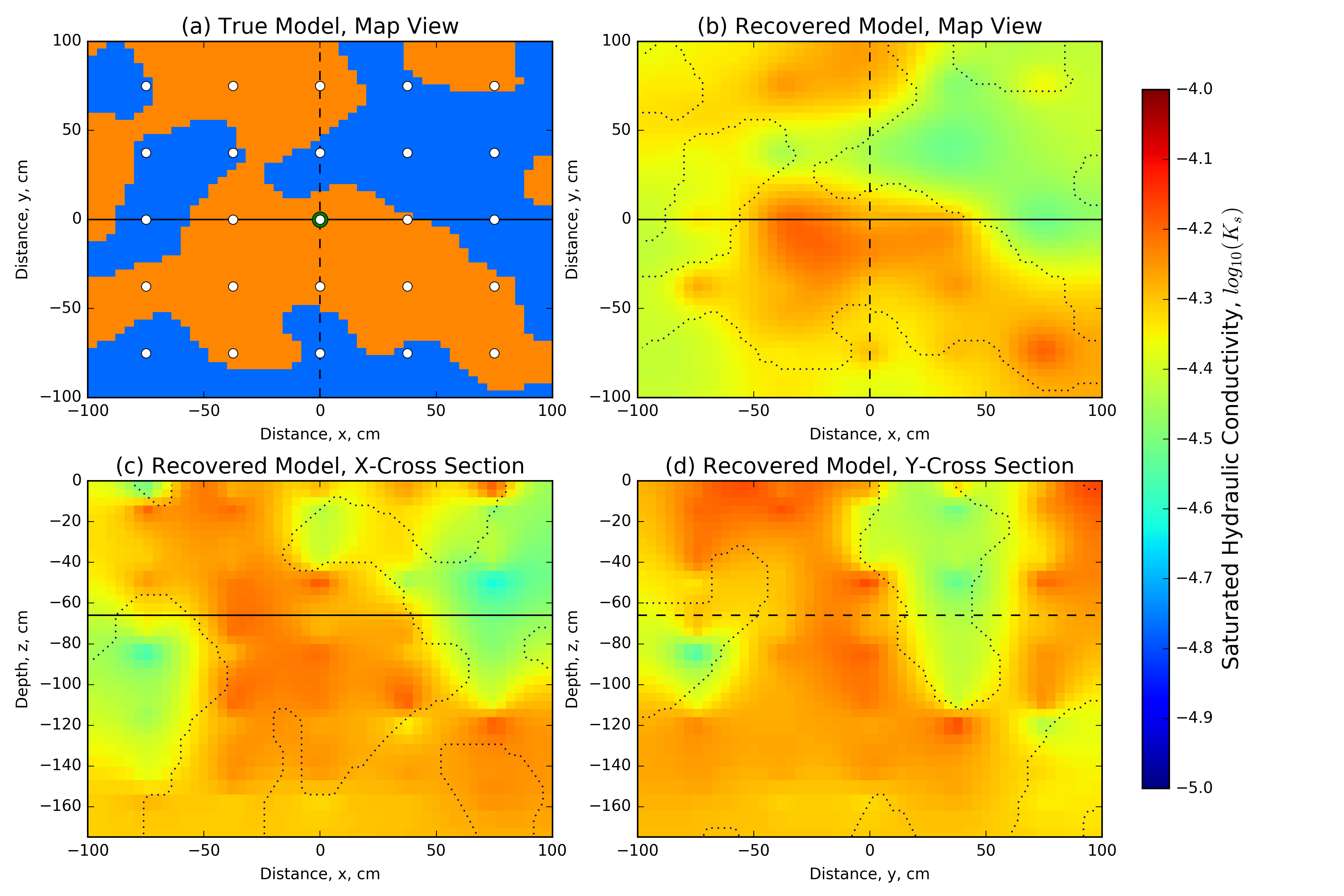}
\end{center}
\caption{The 3D distributed saturated hydraulic conductivity model recovered from the inversion compared to the (a) synthetic model map view section, using (b) the same map view section, (c) an X-Z cross section and (d) a Y-Z cross-section. The synthetic model is show as an outline on all sections, and tie lines are show on all sections as solid and dashed lines, all location measurements are in centimeters.}
\label{fig:richards-inversion3d-results}
\end{figure}

For calculation of synthetic data, the initial conditions are a dry soil with a homogeneous pressure head ($\psi=-30$cm). The boundary conditions applied simulate an infiltration front applied at the top of the model, $\psi = -10$cm $\in \delta\Omega^\text{top}$. Neumann (no-flux) boundary conditions are used on the sides of the model domain.  Figure~\ref{fig:richards-inversion3d-results} shows the pressure head and water content fields from the forward simulation. Figure~\ref{fig:richards-fields}a and \ref{fig:richards-fields}b show two cross-sections at time 5.2 hours and 10.3 hours of the pressure head field. These figures show true soil type model as an outline, where the inclusions are the less hydraulically-conductive loamy sand. The pressure head field is continuous across soil type boundaries and shows the infiltration moving vertically down in the soil column. We can compute the water content field from the pressure head field using the nonlinear van Genuchten model chosen; Figure~\ref{fig:richards-fields}c and \ref{fig:richards-fields}d show this computation at the same times. The loamy sand has a higher relative water content for the same pressure head and the water content field is discontinuous across soil type boundaries, as expected.

\begin{figure}[!htbp]
\begin{center}
\includegraphics[width=0.9\textwidth]{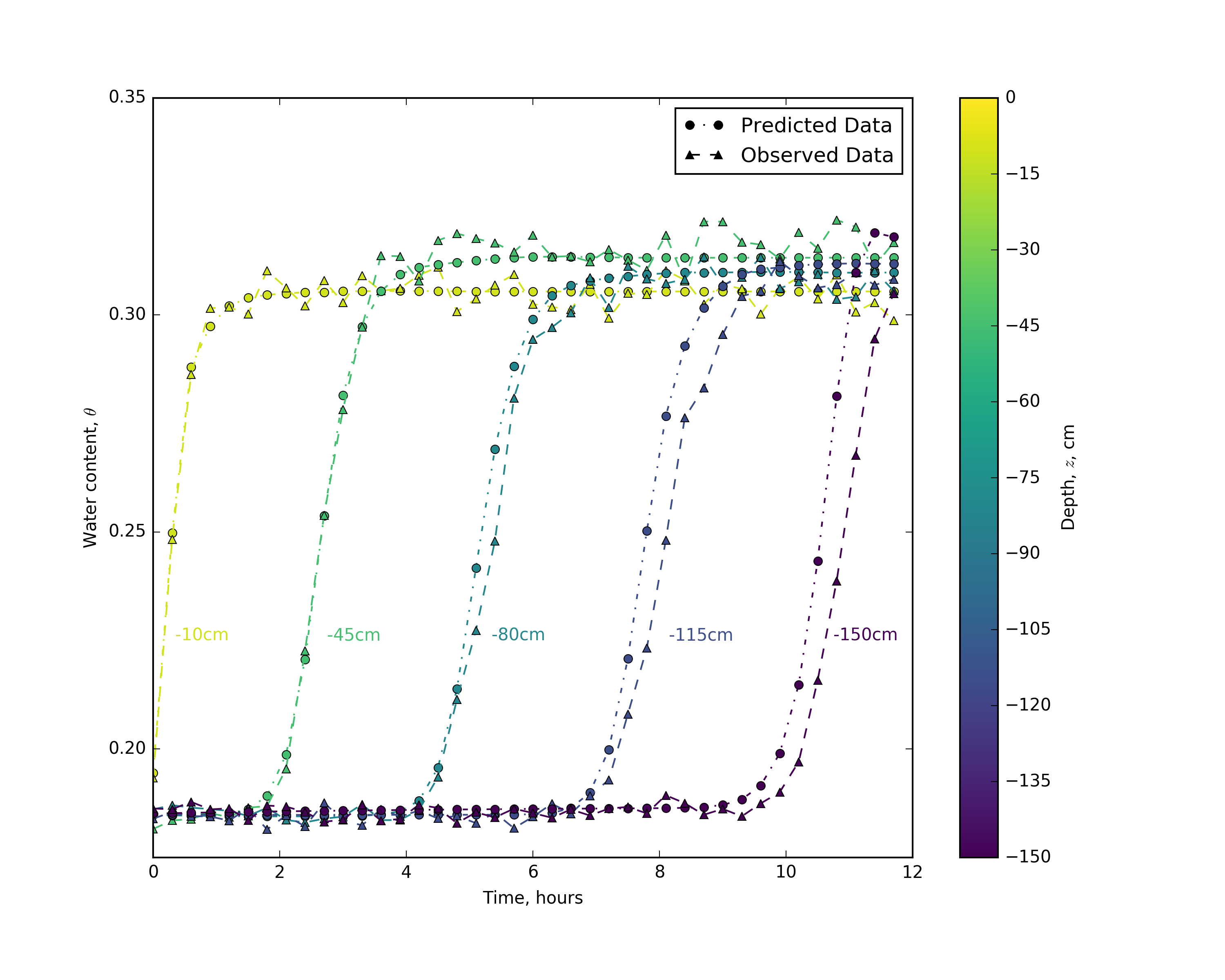}
\end{center}
\caption{Observed and predicted data for five measurements locations at depths from 10cm to 150cm from the center of the model domain.}
\label{fig:richards-data}
\end{figure}

The observed data, which will be used for the inversion, is collected from the water content field at the points indicated in both Figure~\ref{fig:richards-fields}c and \ref{fig:richards-fields}d. The sampling location and density of this three-dimensional grid within the model domain is similar to the resolution of a 3D electrical resistivity survey. Our implementation supports data as either water content or pressure head; however, proxy water content data is more realistic in this context. Similar to the field example in \citep{Pidlisecky2013}, we collect water content data every 18 minutes over the entire simulation, leading to a total of 5000 spatially and temporally extensive measurements. The observed water content data for a single infiltration curve is plotted through depth in Figure~\ref{fig:richards-data}. The green circles in Figure~\ref{fig:richards-fields} show the locations of these water content measurements. The depth of the observation is colour-coded by depth, with the shallow measurements being first to increase in water content over the course of the infiltration experiment. To create the observed dataset, $\bfdo$, from the synthetic water content field, 1\% Gaussian noise is added to the true water content field.  This noise is below what can currently be expected from a proxy geophysical measurement of the water content. However, with the addition of more noise, we must reduce our expectations of our ability to recover the true parameter distributions from the data. In this experiment, we are interested in examining what is possible to recover under the best of circumstances, and therefore have selected a low noise level.

For the inverse problem, we are interested in the distribution of soil types that fits the measured data. We parameterize these soil types using the van Genuchten empirical model (\ref{eq:van-genuchten}) with as least five spatially distributed properties. Inverting for all 5.625$\times10^5$ parameters in this simulation with only 5000 data points is a highly underdetermined problem, and thus there are many possible models that may fit those data. In this 3D example, we will invert solely for saturated hydraulic conductivity and assume that all other van Genuchten parameters are equivalent to the sand; that is, they are parameterized to the \emph{incorrect values} in the loamy sand. Note that this assumption, while reasonable in practice, will handicap the results, as the van Genuchten curves between these two soils differ. Better results can, of course, be obtained if we assume that the van Genuchten parameters are known; this assumption is unrealistic in practice, which means that we will not be able to recreate the data exactly. However, the distribution of saturated hydraulic conductivity may lead to insights about soil distributions in the subsurface. Figure~\ref{fig:richards-inversion3d-results} shows the results of the inversion for saturated hydraulic conductivity as a map view slice at 66 cm depth and two vertical sections through the center of the model domain. The recovered model shows good correlation to the true distribution of hydraulic properties, which is superimposed as a dashed outline. Figure~\ref{fig:richards-data} shows the predicted data overlaid on the true data for five water content measurement points through time; these data are from the center of the model domain and are colour-coded by depth. As seen in Figure~\ref{fig:richards-data}, we do a good job of fitting the majority of the data. However, there is a tendency for the predicted infiltration front to arrive before the observed data, which is especially noticeable at deeper sampling locations. The assumptions put on all other van Genuchten parameters to act as sand, rather than loamy sand, lead to this result.

\subsection{Scalability of the implicit sensitivity}
For the forward simulation presented, the Newton root-finding algorithm took 4-12 iterations to converge to a tolerance of $1\times10^{-4}$m on the pressure head. The inverse problem took 20 iterations of inexact Gauss-Newton with five internal CG iterations used at each iteration. The inversion fell below the target misfit of 5000 at iteration 20 with $\phi_d=4.893\times10^3$; this led to a total of 222 calls to functions to solve the products $\bfJ \bfv$ and $\bfJ^\top \bfz$. Here, we again note that the Jacobian is neither computed nor stored directly, which makes it possible to run this code on modest computational resources; this is not possible if numerical differentiation or direct computation of the Jacobian is used. For these experiments, we used a single Linux Debian Node on Google Compute Engine (Intel Sandy Bridge, 16 vCPU, 14.4 GB memory) to run the simulations and inversion. The forward problem takes approximately 40 minutes to solve. In this simulation, the dense Jacobian matrix would have 562.5 million elements. If we used a finite difference algorithm to explicitly calculate each of the 1.125e5 columns of the Jacobian with a simple forward difference, we would require a calculation for each model parameter -- or approximately 8.5 years of computational time. Furthermore, we would need to recompute the Jacobian at each iteration of the optimization algorithm. In contrast, if we use the implicit sensitivity algorithm presented in this paper, we can solve the entire inverse problem in 34.5 hours.

\begin{table}[!ht]
\centering
\caption{Comparison of the memory necessary for storing the dense explicit sensitivity matrix compared to the peak memory used for the implicit sensitivity calculation excluding the matrix solve. The calculations are completed on a variety of mesh sizes for a single distributed parameter ($K_s$) as well as for five distributed van Genuchten model parameters ($K_s, \alpha, n, \theta_r$, and $\theta_s$). Values are reported in gigabytes (GB).}
\begin{tabular}{*{5}c}
\hline
 &   \multicolumn{2}{c}{\bf Explicit Sensitivity} &    \multicolumn{2}{c}{\bf Implicit Sensitivity}  \\
Mesh Size & 1 parameter & 5 parameters & 1 parameter & 5 parameters  \\
\hline
$32 \times 32 \times 32$    & 1.31 &  6.55 &  0.136 & 0.171\\
$64 \times 64 \times 64$    & 10.5 &  52.4 &  0.522 & 0.772\\
$128 \times 128 \times 128$ & 83.9 &  419  &  3.54  & 4.09 \\
\hline
\end{tabular}
\label{table:richards-sensitivity}
\end{table}



Table~\ref{table:richards-sensitivity} shows the memory required to store the explicit sensitivity matrix for a number of mesh sizes and contrasts them to the memory required to multiply the implicit sensitivity by a vector. These calculations are modifications on the example presented above and use 5000 data points. The memory requirements are calculated for a single distributed parameter ($K_s$) as well as five spatially distributed parameters ($K_s, \alpha, n, \theta_r$, or $\theta_s$). Neither calculation includes the memory required to solve the matrix system, as such, the reported numbers underestimate the actual memory requirements for solving the inverse problem. The aim of this comparison is to demonstrate how the memory requirements scale, an appropriate solver must also be chosen for either method to solve the forward problem. When using an explicitly calculated sensitivity matrix to invert for additional physical properties, the memory footprint increases proportionally to the number of distributed physical properties; this is not the case for the implicit sensitivity calculation. For example, on a $128\times128\times128$ mesh, the explicit formation of the sensitivity requires 419 GB for five spatially distributed model parameters, which is five times the requirement for a single distributed model parameter (83.9 GB). For the implicit sensitivity on the same mesh, only 4.09 GB of memory is required, which is 1.2 times the requirement for a single distributed model parameter (3.54 GB). For this mesh, inverting for five spatially distributed parameters requires over 100 times less memory when using the implicit sensitivity algorithm, allowing these calculations to be run on modest computational resources.

\section{Conclusions}

The number of parameters that are estimated in Richards equation inversions has grown and will continue to grow as time-lapse data and geophysical data integration become standard in site characterizations. The increase in data quantity and quality provides the opportunity to estimate spatially distributed hydraulic parameters from Richards equation; doing so requires efficient simulation and inversion strategies. In this paper, we have shown a computationally efficient derivative-based optimization algorithm that does not store the Jacobian, but rather computes its effect on a vector (i.e. $\bf Jv$ or $\bf J^\top z$). By not storing the Jacobian, the size of the problem that we can invert becomes much larger. We have presented efficient methods to compute the Jacobian that can be used for all empirical hydraulic parameters, even if the functional relationship between parameters is obtained from laboratory experiments.

Our technique allows a deterministic inversion, which includes regularization, to be formulated and solved for any of the empirical parameters within Richards equation. For a full 3D simulation, as many as ten spatially distributed parameters may be needed, resulting in a highly non-unique problem; as such, we may not be able to reasonably estimate all hydraulic parameters. Depending on the setting, amount of a-priori knowledge, quality and quantity of data, the selection of which parameters to invert for may vary. Our methodology enables practitioners to experiment in 1D, 2D and 3D with full simulations and inversions, in order to explore the parameters that are important to a particular dataset. Our numerical implementation is provided in an open source repository (\url{http://simpeg.xyz}).

\section*{Acknowledgments}
The funding for this work is provided through the Vanier Canada Graduate Scholarships Program and grants from The University of British Columbia (NSERC 22R47082). Thank you also to contributors of the \textsc{\textsc{SimPEG}} project (http://simpeg.xyz).

\newpage

\bibliographystyle{iopart-num}
\bibliography{refs.bib}
\end{document}